\documentclass[english]{article}
\usepackage[T1]{fontenc}
\usepackage{babel}
\usepackage[margin=1.0in]{geometry}

\usepackage{amsmath}
\usepackage{algorithm}
\usepackage{algorithmic}
\usepackage{cite}
\usepackage{graphicx}
\usepackage{url}
\RequirePackage[colorlinks,citecolor=blue,urlcolor=blue,linkcolor=blue]{hyperref}
\usepackage{color}

\algsetup{indent=2em}
\newcommand{\algorithmicbreak}{\textbf{break}}
\newcommand{\BREAK}{\STATE \algorithmicbreak}
\begin{document}

\title{Competing Sudakov Veto Algorithms}
\author{Ronald Kleiss \qquad Rob Verheyen \\ \\ \small Institute for Mathematics, Astrophysics and Particle Physics, \\ \small Faculty of Science, Mailbox 79,
Radboud University Nijmegen, \\ \small P.O. Box 9010, 6500 GL Nijmegen, The Netherlands}

\maketitle

\begin{abstract}
We present a formalism to analyze the distribution produced by a Monte Carlo algorithm. We perform these analyses on several versions of the Sudakov veto algorithm, adding a cutoff, a second variable and competition between emission channels. The formal analysis allows us to prove that multiple, seemingly different competition algorithms, including those that are currently implemented in most parton showers, lead to the same result. Finally, we test their performance 
in a semi-realistic setting and show that there are significantly faster alternatives to the commonly used algorithms.
\end{abstract}

\section{Introduction}
Parton showers form an integral part of the event generators that are commonly used to compare data from collider experiments with theory \cite{PYTHIA, SHERPA, HERWIG}. The Sudakov veto algorithm is used in the procedure of generating the subsequent emissions that make up the shower. It facilitates the resummation of logarithmic contributions to all orders in the coupling constant in a Monte Carlo framework, thereby producing realistic final states. 
A positive ordering variable (scale) $t$ is typically evolved down from an initial scale $u$, generating ordered branchings
of partons. The scale of the next branching is selected according
to a probability distribution of the form 

\begin{equation} \label{distr}
E(t;u)=p(t)\Delta(t,u)\mbox{ where }\Delta(t,u) \equiv \exp\left(-\int_{t}^{u}p\left(\tau\right)d\tau\right),
\end{equation}

where the function $p(t)$ is the \emph{branching kernel}. The function $\Delta(t,u)$ is known as the \emph{Sudakov form
factor}. It represents the probability of no emission occurring between two scales. In a Monte Carlo setting, scales must be sampled from eq. \eqref{distr}. To do that, the inverse of the Sudakov form factor must be computed. Unfortunately, $p(t)$ is typically not simple enough for this inverse to be analytically calculable. Therefore, the \emph{Sudakov veto algorithm} is used. In this paper, we will present a thorough analysis of this algorithm. In a practical setting, eq. \eqref{distr} has to be extended in several ways, one of which is the competition between branching channels. We will analyze the veto algorithm for these extensions, and we will in particular provide multiple algorithms to handle competition. Among these algorithms are those used currently by event generators, and some alternatives which, although seemingly different, will be shown to be equivalent. By implementing them in an antenna parton shower much like  \cite{VINCIA1, VINCIA2, VINCIA3}, we test their performance and show that the alternative algorithms are much faster

This paper is organized as follows. In section 2, we will first set up a formalism to analyze Monte Carlo algorithms in general. This formalism is then used to show the validity of the Sudakov veto algorithm in section 3. Next, in section 4, the algorithm is extended to include a cutoff scale, a second variable and competition between branching channels. We will then prove the equivalence of several different algorithms for competition. In section 5, the performance of these algorithms is tested by implementing them in an actual parton shower. 

\section{The unitary algorithm formalism}

A useful approach to the analysis of algorithms can be formulated in terms of integration results.
This can be denoted the formalism of unitary algorithms.
The idea is that these integration results can be translated, at the one hand, into positive statements and, on the other hand, into readily implementable pseudocode. Let $g(x)$ be a probability density.
Then, the formula
\begin{equation}
1 = \int\;g(x)\;dx
\end{equation}

on the one hand reads `we have an algorithm to generate random numbers according to the
distribution $g(x)$', and, on the other hand, the pseudocode statement

\begin{equation}
x \leftarrow g
\end{equation}

which says that the number $x$ be obtained from the algorithm delivering the distribution $g$.
As a simple example, the statement

\begin{equation}
1 = \int_0^1\;dx
\label{onestatement}
\end{equation}

implies that we have available an algorithm that delivers random numbers $x$, uniformly
distributed with the density $\theta(0<x<1)$, where we have defined the logical step function

\begin{equation}
\theta(S) = \left\{\begin{tabular}{cl}1 & if the statement $S$ is true$,$\\
0 & if the statement $S$ is false$.$\end{tabular}\right.
\end{equation}

And indeed, this just says `we generate a random number uniformly distributed between 0 and 1',
using the pseudorandom number generator of choice\footnote{A possible source of conflict is that
the formalism uses the real-number model of computation, while of course the actual code
uses finite-wordsize numbers. On the other hand, {\em any\/} algorithm that is sensitive to the
difference between the two models of computation is tainted and should be shunned.}. In fact, to shorten notation later on, we will denote any random number generated according to eq. \eqref{onestatement} by $\rho$. A second ingredient of the formalism is the assignment operation

\begin{equation}
1 = \int dy\;\delta(y - h(x)),
\end{equation}

which is equivalent to the pseudocode statement
\begin{equation}
y \leftarrow h(x).
\end{equation}

We shall of course use the standard result

\begin{equation}
\delta(y - h(x)) = \sum_j\;\frac{1}{|h'(x)|}\,\delta(x-x_j),
\end{equation}

where the sum runs over the roots $x_j$ of $h(x)=y$ (all assumed to be single). It is to be noted here that the integral over $y$ runs over all real values, but if the range of
$h$ is restricted to $h_0 \le h(x) \le h_1$, then we automatically have the corresponding bounds on $y$.

As a simple example, let us imagine the inverse of the primitive function $P(t)$ of $p(t)$ from eq. \eqref{distr} is available. The pseudocode to generate values of $t$ according to eq. \eqref{distr} is:

\begin{equation} \label{newtstep}
t \leftarrow P^{-1}\left(\log\left(\rho\right)+P\left(u\right)\right),
\end{equation}

where $\rho$ here and in the following comes from an (idealized) source of iid\footnote{independent, identically distributed.} random numbers uniform in $(0,1]$. We analyze eq. \eqref{newtstep}:
\begin{align} \label{algStep}
1 = &\int_{0}^{1}d\rho\int dt\,\delta\left(t-P^{-1}\left(\log\left(\rho\right)+P\left(u\right)\right)\right) \nonumber \\
   = &\int_{0}^{1}d\rho\int dt\,\delta\left(P\left(t\right)-\log\left(\rho\right)-P\left(u\right)\right)p\left(t\right) \nonumber \\
   = &\int_{0}^{1}d\rho\int dt\,\delta\left(\rho-e^{P\left(t\right)-P\left(u\right)}\right)e^{P(t)-P(u)}p(t) \nonumber \\
   = &\int_{0}^{u}dt\,p(t)e^{-\int_{t}^{u}p(t)}.
\end{align}

so that `we have an algorithm to generate $t$ according to eq. \eqref{distr}', where the algorithm is of course given by eq. \eqref{newtstep}.
 
A variant of the formalism is encountered in the rejection algorithm, which is already very close to the Sudakov veto algorithm. Let $g(x)$ be a probability density that we {\em can\/} generate,  $f(x)$ a non-negative function, and $c$ a number such that $c\,g(x)\ge f(x)$ over the support of $f(x)$. The rejection algorithm then reads

\begin{algorithm}[H]
\caption{The rejection algorithm}
\begin{algorithmic}
\LOOP 
\STATE x $\leftarrow$ g
\IF {$c\,\rho \le f(x)/g(x)$}
\RETURN $x$
\ENDIF
\ENDLOOP
\end{algorithmic}
\end{algorithm}

Let $K(x)$ be the resulting density. We can then write

\begin{align}
\label{rejectionalgorithm}
K(x) &= \int dy\;g(y)\;\int_0^1 d\rho\;\left[
\theta\left(\rho\le \frac{f(y)}{c\,g(y)}\right)\,\delta(x-y) + 
\theta\left(\rho> \frac{f(y)}{c\,g(y)}\right)\,K(x)\right] \nonumber \\
&= \int dy\;g(y)\;\left[ \frac{f(y)}{c\,g(y)}\,\delta(x-y) + \left(1- \frac{f(y)}{c\,g(y)}\right)\,K(x)\right] \nonumber \\
&= \int dy\;\left[\frac{f(y)}{c}\, \delta(x-y) + g(y)\,K(x) - \frac{f(y)}{c}\,K(x)\right] \\ \nonumber
&= \frac{1}{c}f(x) + K(x) - \frac{1}{c}\int dy\;f(y)\,K(x),
\end{align}

from which we see that $K(x)$ is the normalized probability density proportional to $f(x)$:

\begin{equation}
K(x) = \frac{f(x)}{\int dy f(y)}.
\end{equation}
Note how the loop is embodied by the reappearance of $K(x)$ on
the right-hand side in the first line of eq. \eqref{rejectionalgorithm}. With these few basic
ingredients the result of any algorithm (provided it terminates with unit probability) can be
reduced to the elimination of Dirac delta functions, and we shall employ these ideas in what follows.

\section{Analyzing the Sudakov veto algorithm}
We now present the Sudakov veto algorithm and analyze it using the techniques of the previous section. We first establish that eq. \eqref{distr} is normalized if $P(t)$, the primitive
function of $p(t)$, goes to $-\infty$ as $t\rightarrow 0$:

\begin{equation}
E\left(t;u\right)=\frac{\partial}{\partial t}\Delta(t,u)\mbox{ }\rightarrow\mbox{ }\int_{0}^{u}E\left(t;u\right)dt=1-\exp\left(P\left(0\right)-P\left(u\right)\right).
\end{equation}
 
The Sudakov veto algorithm relies on the existence of an overestimate function $q(t) \geq p(t)$ which does have an invertible Sudakov factor. The algorithm is given below in pseudocode.
\begin{algorithm}[H]
\caption{The Sudakov veto algorithm}
\label{simpleVeto}
\begin{algorithmic}
\STATE $t \leftarrow u$
\LOOP 
\STATE $t \leftarrow Q^{-1}\left(\log\left(\rho_1\right)+Q\left(t\right)\right)$
\IF {$\rho_{2}<p(t)/q(t)$}
\RETURN $t$
\ENDIF
\ENDLOOP
\end{algorithmic}
\end{algorithm}

It was shown in the previous section that the first step in the loop generates values of $t$ distributed according to eq. \eqref{distr} where the kernel is $q(t)$ instead of $p(t)$, and the scale $u$ is set to the previous value of $t$. Thus, the value of $t$ is evolved downward at every step of the loop, which is the crucial difference with algorithm \ref{rejectionalgorithm}. There, subsequent values for $t$ would be generated in the same way every time. The if-statement represents the veto step. A scale is accepted with probability $p(t)/q(t)$, at which point the algorithm terminates. We now convert the algorithm to unitary language as we did before in eq. \eqref{rejectionalgorithm} for the rejection algorithm.

\begin{align}
E(t;u) &= \int_{0}^{u} dt\int_{0}^{u}d\tau\,q\left(\tau\right)e^{Q\left(\tau\right)-Q\left(u\right)}  \nonumber \\
&\times \int_{0}^{1}d\rho \bigg[ \bigg. \theta\left(\rho<\frac{p(\tau)}{q(\tau)}\right)\delta\left(\tau-t\right) + \theta\left(\rho>\frac{p(\tau)}{q(\tau)}\right)E(t;\tau)\bigg. \bigg].
\end{align}

After generating a trial scale $\tau$, the random number $\rho$ and the step functions guide the algorithm to either accept the generated scale, or to start over using $\tau$ as the new starting point. 
Next, the integral over $\rho$ is worked out.

\begin{equation}
e^{Q(u)}E(t;u)=\int_{0}^{u}d\tau\, e^{Q(\tau)}\left[p(\tau)\delta(t-\tau)+\left(q(\tau)-p(\tau)\right)E(t;\tau)\right].
\end{equation}

Taking the derivative with respect to $u$, we find the following differential equation:

\begin{equation}
\label{difEq}
\frac{\partial}{\partial u}E(t;u)=p(u)\delta(t-u)-p(u)E(t;u).
\end{equation}

It is solved by

\begin{equation}
\label{regDensity}
E(t;u)=p(t)\exp\left(-\int_{t}^{u}dx\,p(x)\right)\theta\left(0<t<u)\right),
\end{equation}

which is eq. \eqref{distr}. It is, however, not the most general solution to eq. \eqref{difEq}. We will consider this issue more carefully in the next section.

\section{Extending the algorithm}

Next, we consider the Sudakov veto algorithm in a more practical setting. The algorithm needs to be extended in several ways to be applicable in a real parton shower. They are:
\begin{itemize}
\item An infrared cutoff $\mu$ is has to be introduced. This cutoff is required in QCD to avoid the nonperturbative regime. In event generators, the parton shower is evolved to this cutoff scale, after which the results are fed to a hadronization model. The consequence is that the Sudakov factor will not equal zero at the lower boundary of the scale integral. Therefore eq. \eqref{distr} is no longer normalized to one and is thus not a probability distribution.
\item The scale variable $t$ is not enough to parameterize the entire branching phase space. An additional variable $z$ has to be introduced \footnote{Actually, a third parameter is required. This is usually taken to be the azimuthal angle $\phi$. We will assume $\phi$-independent branching kernels, such that the $\phi$ integral is trivial.}. In traditional parton showers, this parameter is the energy fraction carried by a newly created parton. However, in the more modern dipole or antenna showers, it is just a variable that parameterizes the factorized phase space. The boundaries of the branching phase space translate to scale-dependent boundaries on $z$.
\item The algorithm has to account for emissions from multiple channels. These channels can originate from either the presence of multiple partons or dipoles, or from multiple branching modes.  
\end{itemize}
We now include these issues separately before incorporating them into a single algorithm.

\subsection{Introducing a cutoff}
In a realistic parton shower, the values of the scale $t$ are not allowed to reach zero. In the case of QCD, a cutoff value $\mu$ is set at a value of about $1$ GeV, below which a perturbative approach is no longer valid.
Eq. \eqref{distr} now no longer represents a probability distribution. This same problem would occur if the primitive of the branching kernel $P(t)$ would not diverge for vanishing $t$, as is for instance the case for kernels of massive particles. The following algorithm, due to \cite{Reloaded}, allows for the introduction of a cutoff and deals with non-diverging $P(t)$ simultaneously. The algorithm below first shows how to generate trial values for $t$.

\begin{algorithm}[H]
\caption{Generate trial scales in the presence of a cutoff $\mu$}
\label{cutTrialA}
\begin{algorithmic}
\IF {$\rho < \rho_{c} = e^{Q(\mu) - Q(t_0)}$}
\STATE $t \leftarrow \mu$
\ELSE
\STATE $t \leftarrow Q^{-1}(\mbox{log}(\rho) + Q(t_0))$
\ENDIF
\RETURN $t$
\end{algorithmic}
\end{algorithm}

We analyze this algorithm to find what probability distribution it represents.
\begin{align}
\label{cutTrialE}
\bar{E}(t;\mu,u) &=\int_{0}^{1}d\rho\left[\theta\left(\rho\leq\rho_{c}\right)\delta\left(t-\mu\right)+\theta\left(\rho>\rho_{c}\right)\delta\left(t-Q^{-1}\left(\log(\rho)+Q(t_0)\right)\right)\right] \nonumber \\
&=\rho_{c}\,\delta\left(t-\mu\right)+\int_{\rho_{c}}^{1}d\rho \,q(t)e^{Q(t)-Q(t_0)}\delta\left(\rho-e^{Q(t)-Q(t_0)}\right) \nonumber \\
&=\rho_{c}\,\delta\left(t-\mu\right)+q(t)e^{Q(t)-Q(t_0)}\theta\left(\rho_{c}<e^{Q(t)-Q(t_0)}<1\right) \nonumber \\
&=e^{Q(\mu)-Q(t_0)}\delta\left(t-\mu\right)+q(t)e^{Q(t)-Q(t_0)}\theta(\mu<t<t_0).
\end{align}

where in the last step we used the fact that $q(t)$ is a positive function, and therefore $Q(t)$ is monotonically increasing. Compared with eq. \eqref{algStep}, eq. \eqref{cutTrialE} has an additional term that compensates the contribution of the lower bound on the original probability distribution. The veto algorithm should reproduce this distribution for the branching kernel $p(t)$. 

\begin{algorithm}[H]
\caption{The Sudakov veto algorithm in the presence of a cutoff $\mu$}
\begin{algorithmic}
\STATE $t \leftarrow u$
\LOOP 
\STATE $t \leftarrow$ Algorithm \ref{cutTrialA}
\IF{$t = \mu$}
\RETURN $\mu$
\ELSIF {$\rho_{2}<p(t)/q(t)$} 
\RETURN $t$
\ENDIF 
\ENDLOOP
\end{algorithmic}
\end{algorithm}

Writing it down in unitary language:
\begin{align}
E(t;u) &=\int d\tau\left(e^{Q(\mu)-Q(u)}\delta(\tau-\mu)+q(\tau)e^{Q(\tau)-Q(u)}\theta(\mu<\tau<u)\right) \nonumber \\
&\times \Biggl\{ \theta\left(\tau=\mu\right)\delta\left(t-\mu\right)+ \theta\left(\tau\neq\mu\right)\Biggr. \nonumber \\ 
&\times \Biggr.\int_{0}^{1}d\rho\left[\theta\left(\rho<\frac{p(\tau)}{q(\tau)}\right)\delta\left(t-\tau\right)+\theta\left(\rho>\frac{p(\tau)}{q(\tau)}\right)E(t;\tau)\right] \Biggr\}.
\end{align}
Going through the same steps as before, we find
\begin{align}
\label{cutVeto}
e^{Q(u)}E(t;u)&=e^{Q(\mu)}\delta\left(t-\mu\right) \nonumber \\
&+\int_{\mu}^{u}d\tau\, e^{Q(\tau)}\left[p(\tau)\delta(t-\tau)+\left(q(\tau)-p(\tau)\right)E(t;\tau)\right].
\end{align}
After taking the derivative with respect to u, the first term drops out and the $\mu$-dependence disappears from the second. Therefore, eq. \eqref{difEq} is recovered. However, eq. \eqref{regDensity} is not the only solution to this differential equation. A more general solution is:

\begin{equation}
E(t;u) = e^{P(\sigma)-P(u)}\delta\left(t-\sigma\right)+p(t)e^{P(t)-P(u)}\theta(\sigma<t<u)
\end{equation}

for some scale $\sigma < u$. To fix sigma, we require that $E(t;u)$ reduces to a delta function distribution when $u \rightarrow \mu$, which leads to $\sigma = \mu$.

\subsection{Introducing a second variable}
The targeted distribution is now:
\begin{equation} \label{bivar}
E(t,z;u)=p(t,z)\Delta(u,t)\mbox{ where }\Delta(u,t)=\exp\left(-\int_{t}^{u}d\tau\int_{z_{-}(\tau)}^{z_{+}(\tau)}d\zeta\,p(\tau,\zeta)\right),
\end{equation}
which is normalized as
\begin{equation}
\int_{0}^{u}dt\int_{z_{-}(t)}^{z_{+}(t)}dz\,E(t,z;u)=1.
\end{equation}
We now need to produce pairs $(t,z)$ distributed according to $E(t,z;u)$. A difficulty lies in the dependence of the range of $z$ on the scale. In order to generate a value for $t$, the $\zeta$ integral in the Sudakov factor is required, which depends on $t$. On the other hand, $z$ cannot be generated first, since its boundaries depend on $t$. 

To deal with this problem, an additional veto condition is introduced. We introduce a constant overestimate of the $z$-range as $z_{-} \leq z_{-}(t)$ and $z_{+} \geq z_{+}(t)$. Additionally we require the overestimate function to be factorized as $q(t,z)=r(t)s(z)$ where still $q(t,z) \geq p(t,z)$. Then, we define 

\begin{equation}
q(t) \equiv r(t)\int_{z_{-}}^{z_{+}}dz\,s(z) = r(t)\left(S(z_{+})-S(z_{-})\right).
\end{equation}
The algorithm is given below.

\begin{algorithm}[H]
\caption{The Sudakov veto algorithm for two variables} \label{twoVar}
\begin{algorithmic}
\STATE $t \leftarrow u$
\LOOP 
\STATE $t \leftarrow Q^{-1}\left(\log\left(\rho_{1}\right)+Q\left(t\right)\right)$
\STATE $z \leftarrow S^{-1}\left(\rho_{2}\left(S(z_{+})-S(z_{-})\right)+S(z_{-})\right)$
\IF {$\rho_{3}<p(t,z)/q(t,z)$ and $z_-(t) < z < z_+(t)$}
\RETURN $t$
\ENDIF 
\ENDLOOP
\end{algorithmic}
\end{algorithm}

We first analyze the step of this algorithm that generates $z$.

\begin{align}
1= &\int_{0}^{1}d\rho_{2}\int dz\,\delta\left(z-S^{-1}\left[\rho_{2}\left(S(z_{+})-S(z_{-})\right)+S(z_{-})\right]\right) \nonumber \\
= &\int_{0}^{1}d\rho_{2}\int dz\,\delta\left(S(z)-\rho_{2}\left(S(z_{+})-S(z_{-})\right)+S(z_{-})\right)s(z) \nonumber \\
  = &\int_{z_{-}}^{z_{+}}dz\,\frac{s(z)}{S(z_{+})-S(z_{-})}.
\end{align}

Thus, $z$ is distributed according to $s(z)$. Introducing the notation 

\begin{equation}
\theta^{\tau}(\zeta) \equiv \theta(z_{-}(\tau) < \zeta < z_{+}(\tau)),
\end{equation}

we now analyze algorithm \ref{twoVar}.

\begin{align}
E\left(t,z;u\right) &= \int_{0}^{u}q(\tau)e^{Q(\tau)-Q(u)}\int_{z_{-}}^{z_{+}}d\zeta\frac{s(\zeta)}{S(z_{+})-S(z_{-})}\int_{0}^{1}d\rho\,  \nonumber \\
&\times \Bigl\{ \Bigr. \left(1-\theta^{\tau}(\zeta)\right)E\left(t,z;\tau\right) +\theta^{\tau}(\zeta)\theta\left(\rho>\frac{p(\tau,\zeta)}{q(\tau,\zeta)}\right)E\left(t,z;\tau\right)  \nonumber \\
&+\Bigl.\theta^{\tau}(\zeta)\theta\left(\rho<\frac{p(\tau,\zeta)}{q(\tau,\zeta)}\right)\delta\left(\tau-t\right)\delta\left(\zeta-z\right)\Bigr\}. 
\end{align}

Evaluating the integrals and taking the derivative with respect to $u$ leads to:

\begin{equation}
\frac{\partial}{\partial u}E\left(t,z;u\right)=p(u,z)\delta(u-t)\theta_{z}-\int_{z_{-}(t)}^{z_{+}(t)}d\zeta\,p(u,\zeta)E\left(t,z;u\right),
\end{equation}

which is solved by eq. \eqref{bivar}.
\subsection{Competing channels}
Let us assume there are $n$ branching channels, each characterized by a branching kernel $p_{i}(t)$. The density $E(t;u)$ now contains a Sudakov factor representing the no-branching probability for all channels, which is just the product of the individual Sudakov factors. The probability of branching at some scale is the sum of the kernels. Introducing the notation  

\begin{equation}
\widetilde{f}(t) \equiv \sum_{i=1}^{n}f_{i}(t)
\end{equation}

for any set of $n$ functions, this leads to the probability distribution

\begin{equation}\label{compDis}
E(t;u)=\widetilde{p}(t)\Delta(t,u)\mbox{ where }
\Delta(t,u)=\exp\left(-\int_{t}^{u}\widetilde{p}\left(\tau\right)d\tau\right).
\end{equation}

This distribution can be produced by generating multiple scales and selecting the highest. This can be shown using the following result:

\begin{align} \label{oldComp}
1&=\int_{0}^{u}dt\left[\prod_{i=1}^{n}\int_{0}^{u}d\tau_{i}f_{i}(\tau_{i})\exp(F_{i}(\tau_{i})-F_{i}(u))\right] \sum_{j=1}^{n}\theta(\max(\tau_{j}))\,
\delta\left(t-\tau_{j}\right) \nonumber \\
&=\int_{0}^{u}dt\sum_{i=1}^{n}\left[\prod_{j\neq i}\int_{0}^{\tau_{i}}d\tau_{j}f(\tau_{j})\exp(F_{j}(\tau_{j})-F_{j}(u)) \right]\int_{0}^{u}d\tau_{i}f_{i}(\tau_{i})\exp(F_{i}(\tau_{i})-F_{i}(u))\,\delta\left(t-\tau_{i}\right) \nonumber \\
&=\int_{0}^{u}dt\sum_{i=1}^{n}f_{i}(t)\exp(F_{i}(t)-F_{i}(u))\left[\prod_{j\neq i}\exp(F_{j}(t)-F_{j}(u))\right] \nonumber \\ 
&=\int_{0}^{u}dt\,\widetilde{f}(t)\exp(\widetilde{F}(t)-\widetilde{F}(u)),
\end{align}

where we used the notation
\begin{equation}
\theta(\max(\tau_{j})) \equiv \prod_{k\neq j} \theta\left(\tau_{j}>\tau_{k}\right),
\end{equation}
which is a step function selecting the highest of all $\tau$. The functions $f_i$ can be either $p_i$ or $q_i$. In the first case, the veto algorithm for a single channel can be used to produce the densities that appear in the first line of eq. \eqref{oldComp}. In the second case, the highest of the trial scales is selected and subsequently the veto step is applied using the kernel of the selected channel. Both procedures result in eq. \eqref{compDis}. 

Next, we present a very different algorithm that also produces this density. 

\begin{algorithm}[H]
\caption{A different competition Sudakov veto algorithm}
\label{myAlg}
\begin{algorithmic}
\STATE $t \leftarrow u$
\LOOP 
\STATE $t \leftarrow \widetilde{Q}^{-1}\left(\log\left(\rho_{1}\right)+
\widetilde{Q}\left(t\right)\right)$
\STATE Select $i$ between $1$ and $n$ with probability $q_{i}(t)/\widetilde{q}(t)$
\IF {$\rho_{2}<p_{i}(t)/q_{i}(t)$}
\RETURN $t$
\ENDIF 
\ENDLOOP
\end{algorithmic}
\end{algorithm}

We analyze this algorithm to show that it also produces eq. \eqref{compDis}:

\begin{align}
E(t;u)&=\int_{0}^{u}d\tau\,\widetilde{q}(\tau)e^{\widetilde{Q}(\tau)-\widetilde{Q}(u)}\int_{0}^{1}d\rho_{1}\sum_{i=1}^{n}\theta\left(\frac{\sum_{j=0}^{i-1}q_{j}(\tau)}{\widetilde{q}(\tau)}<\rho_{1}<\frac{\sum_{j=0}^{i}q_{j}(\tau)}{\widetilde{q}(\tau)}\right) \nonumber \\
&\times \int_{0}^{1}d\rho_{2}\left[\theta\left(\rho_{2}<\frac{p_{i}(\tau)}{q_{i}(\tau)}\right)\delta\left(t-\tau\right)+\theta\left(\rho_{2}>\frac{p_{i}(\tau)}{q_{i}(\tau)}\right)E(t;\tau)\right],
\end{align}

where $q_{0}(t) \equiv 0$. We go through the usual steps, noting that after doing the $\rho_{1}$ integral, the new sum over step functions yields terms $q_{i}(\tau)/\widetilde{q}(\tau)$ representing the probabilities to select the corresponding channels. The differential equation becomes:
\begin{equation}
\frac{\partial}{\partial u}E(t;u)=\widetilde{p}(u)\delta(t-u)-\widetilde{p}(u)E(t;u),
\end{equation}
which is solved by eq. \eqref{compDis}.

Algorithm \ref{myAlg} requires the generation of trial scales using $\widetilde{q}(t)$ as overestimated branching kernel. In practice, this is often not much harder than generating trial scales for individual channels, since the kernels $q_{i}(t)$ can usually be chosen to have the same $t$-dependence. In such a case, the channel selection step in algorithm \ref{myAlg} does not even require the evaluation of the kernels at the trial scale anymore. We note that algorithm \ref{myAlg} can still be used in more complicated situations by using the procedure outlined in eq. \eqref{oldComp} to split $\widetilde{q}(t)$ up into groups of similar channels. 
In the next chapter, we incorporate the extensions discussed here into a full, practical veto algorithm. Since it was found there are multiple ways to handle competition, these algorithms are tested for their computing times. 

\section{Testing the algorithms}
We now combine all the pieces discussed in the previous section into a single algorithm. Here, we give a description of the full algorithms that all handle competition differently. A concrete statement of the algorithms can be found in the appendix. Additionally, the expression of every algorithm in unitary language is included. These equation can all be shown to be satisfied by:

\begin{align}
E(t,z;u) &= \delta(t-\mu)\delta(z-z_0) \exp\left(-\sum_{i=1}^{n}\int_{\mu}^{u} d\tau \int_{z_{i-}(\tau)}^{z_{i+}(\tau)}d\zeta\,p_i(\tau,\zeta)\right) \\ \nonumber
&+ \sum_{i=1}^{n}f(t,z)\theta_i^t(z)\theta(\mu<t<u) \exp\left(-\sum_{i=1}^{n}\int_{t}^{u} d\tau \int_{z_{i-}(\tau)}^{z_{i+}(\tau)}d\zeta\,p_i(\tau,\zeta)\right).
\end{align}

\begin{itemize}
\item \emph{Veto-Max}: This algorithm handles competition using eq. \eqref{oldComp}, where $f_i(t,z) = p_i(t,z)$. That is, the veto algorithm is applied to every channel individually, then the highest of the generated scales is selected. This is the most common way of handling competition. It is usually cited in the literature as \emph{the} competition algorithm \cite{Fooling, Reloaded}, and is used in most parton showers.
\item \emph{Max-Veto}: This algorithm also uses eq. \eqref{oldComp}, but with $f_i(t,z) = q_i(t,z)$. That is, trial pairs $(t,z)$. The highest of these scales is selected, to which the veto step is applied using the branching kernel of the selected channel. This algorithm is used in the Vincia parton shower \cite{VINCIA1, VINCIA2}.
\item \emph{Generate-Select}: This is the new algorithm described in section 4.3. It generates trial scales $\tau$ using the sum of the overestimate functions $\widetilde{q}(t,z)$. The overestimate functions are required to have the same $z$-dependence. That way, a corresponding $\zeta$ can be generated using boundaries that are overestimates for all channels. Next, a channel $i$ is selected with probability $q_i(\tau)/\widetilde{q}(\tau)$. Then, the veto step is applied to this channel.
\item \emph{Select-Generate}: Under certain circumstances, a slight variation of the Generate-Select algorithm is possible. If we require all overestimate functions $q_i(t,z)$ to have the same scale dependence, this dependence drops out of the selection probabilities. In that case, a channel can be selected before a scale is generated. As a consequence, the overestimate functions can have different dependence on $z$, and universal overestimates are no longer required. 
\end{itemize}  

We test these algorithms by implementing them in a relatively simple antenna shower very close to what is described in \cite{VINCIA1, VINCIA2}. This shower handles QCD radiation using an antenna scheme to include collinear and soft enhancements. It features exact $2 \rightarrow 3$ kinematics for massive particles, but does not include any matching scheme and concerns only final state radiation. It is very basic compared with the parton showers of \cite{PYTHIA, SHERPA, HERWIG} or recent versions of the Vincia shower \cite{VINCIA3}, including only the absolute necessities for a functional parton shower.

The running coupling is taken into account by an overestimate 
\begin{equation}
\hat{\alpha}_s(t) = a\ln^{-1}(bt) 
\end{equation}

where $a$ and $b$ are chosen such that, at the starting scale and the cutoff scale, $\hat{\alpha}_s(t)$ matches the real one-loop running $\alpha_s(t)$, which includes the proper flavor thresholds. This overestimate is corrected by using $\hat{\alpha}_s(t)$ for the overestimate kernels and $\alpha_s(t)$ for the branching kernels.

The possible branchings for a QCD shower can be divided into two categories: emissions, where a quark or gluon sends out a new gluon, and splittings, where a gluon splits into a quark-antiquark pair. We use $p_\perp$-ordering for both for easy application of the Generate-Select and Select-Generate algorithms. The overestimates of the branching kernels are:

\begin{equation}
q_{\mbox{emit}}(t,z) = \frac{2a\,C_{A}}{4\pi \sqrt{\lambda(1,\frac{m_1^2}{s_{12}},\frac{m_2^2}{s_{12}})}} \frac{1}{z(1-z)} \frac{1}{t \ln(bt)} \qquad q_{\mbox{split}}(t,z) = \frac{2a\,n_{F}T_{R}}{4\pi \sqrt{\lambda(1,\frac{m_1^2}{s_{12}},\frac{m_2^2}{s_{12}})}} \frac{1}{z(1-z)} \frac{1}{t \ln(bt)},
\end{equation}

where $\lambda$ is the K{\"a}ll{\'e}n function, $m_1$ and $m_2$ are the masses of the particles in the antenna and $s_{12}$ is its invariant mass. Note that a factor $n_F$ is included in the overestimate of the splitting kernel. It is there because Vincia uses a mix of the Max-Veto and the Generate-Select algorithms. If a gluon splitting is selected through the Max-Veto algorithm, a quark flavor is chosen at random as is done by the Generate-Select algorithm. We use the antennae functions given in given in \cite{VINCIA2} for the splitting kernels. The code can be found in \cite{Code}. 

We compare the performance of the algorithms described above on this shower. In the Veto-Max algorithm we have implemented the following shortcut. While running the single-channel veto algorithm on all available channels, the algorithm keeps track of the highest scale generated thus far. Then, if a scale lower than this highest scale is ever reached, the veto algorithm on the current channel can immediately be aborted. This trick is not available for the Max-Veto algorithm, because it performs the veto step after selecting the highest trial scale between all channels. 

For the Select-Generate algorithm, the bottleneck is the channel selection step. It is complicated by the fact that the K{\"a}ll{\'e}n function and the $z$ integral in the overestimates are different for every antenna. We use stochastic roulette-wheel selection\cite{Roulette} for the selection step, which achieves $\mathcal{O}(1)$ complexity\footnote{Coincidentally, this is also a veto algorithm and is easily provable using unitary language.}. The Generate-Select algorithm assigns the same boundaries for the $z$ integral for all channels, but retains differences in the K{\"a}ll{\'e}n function. We move this difference to the veto step by using the lowest K{\"a}ll{\'e}n function of all antennae for all channels, increasing the overestimation of the branching kernels. Then, for $n_F=6$ and the standard values $C_A = 3$ and $T_R = 1/2$, all overestimate functions are the same, and the channel selection step is trivial. In this sense, the difference between the Generate-Select and the Select-Generate algorithms is a trade-off between easier selection of a channel and lower veto rates.

A remark is in order here. In the splitting $g \to q\, \bar{q}$ the original colour structure is 
separated into two pieces which can be evolved independently. Since our interest here is in the speed of the
various algorithms rather than the development of a fully realistic parton shower, we have not implemented this effect.

We produce $8$ million events per algorithm. The initial scale is $(7\,\mbox{TeV})^2$ and the cutoff scale is $(1\,\mbox{GeV})^2$. These settings produce events with parton multiplicities of $\mathcal{O}(100)$, which are typical at the LHC. To check the equivalence of the veto algorithms, we compute the average amounts of quarks and gluons generated per event. These numbers are very sensitive to small differences in distribution. Table \ref{multtable} shows these averages for every algorithm.

\begin{table}[H]
\centering
\begin{tabular}{|l|l|l|}
\hline
                & Quark Mutliplicity     & Gluon Multiplicity     \\ \hline
Generate-Select & 11.7329 $\pm$ 0.001908 & 64.7354 $\pm$ 0.008516 \\ \hline
Select-Generate & 11.7297 $\pm$ 0.001908 & 64.7359 $\pm$ 0.008514 \\ \hline
Veto-Max        & 11.7294 $\pm$ 0.001907 & 64.7372 $\pm$ 0.008515 \\ \hline
Max-Veto        & 11.7326 $\pm$ 0.001909 & 64.7336 $\pm$ 0.008513 \\ \hline
\end{tabular}
\caption{The average multiplicities produced by the shower with starting at $(7\,\mbox{TeV})^2$ for all veto algorithms.}
\label{multtable}
\end{table}

Figure 1 shows the average amount of CPU time the shower requires to produce events, plotted as a function of the amount of available branching channels as the shower terminates. This measure gives us a good idea of the performance of the algorithms in a practical context. The shape of the curves of the Veto-Max and the Max-Veto algorithms should not be heavily influenced by the specifics of the shower, since factors like branching kernel evaluation times and veto probabilities should be similar for different implementations. However, the relative performance of the Generate-Select and the Select-Generate algorithms does depend on the specific implementation. In this case, the algorithms perform similarly, but this may not be the case for other branching kernels. Either way, the Generate-Select and the Select-Generate algorithms perform much better than the Veto-Max and the Max-Veto algorithms. 

\begin{figure}[H]
\centering
    \input{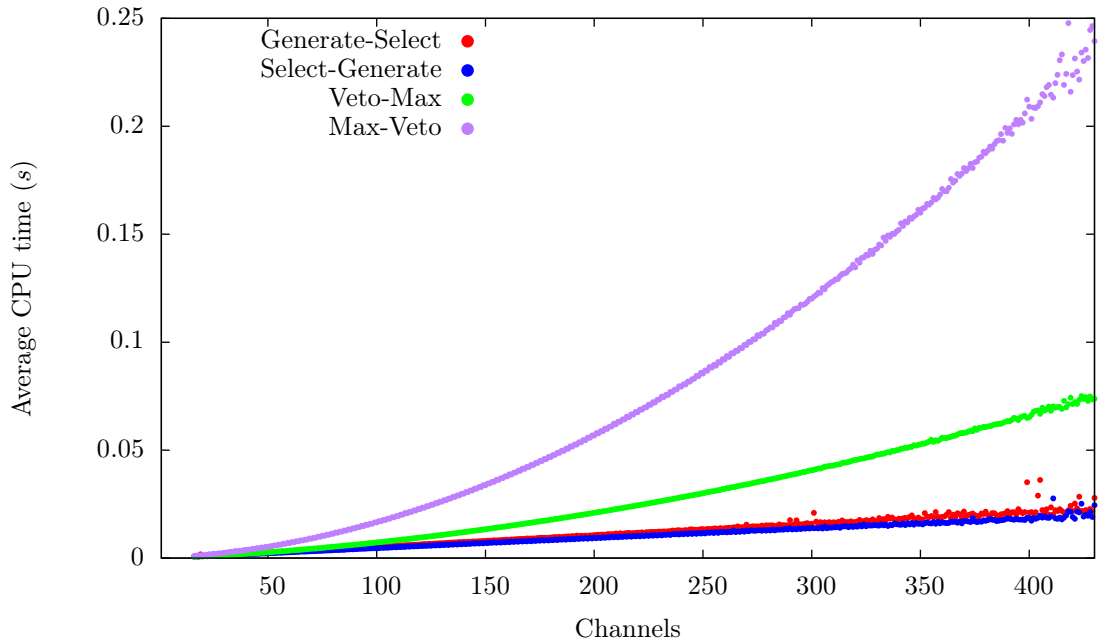}
\caption{The average CPU times required by the shower to produce events as a function of the available branching channels at termination.}
\end{figure}

\section{Conclusion}
The Sudakov veto algorithm forms an integral part of all modern parton shower programs. We describe a formalism that can be used to analyze the distributions that are produced by different versions of this algorithm. Using this method, we discuss various ways of handling competition. While seemingly different, our formal analysis shows that they produce the same distributions. The algorithms were tested using a simple antenna shower, which showed that the new algorithms are faster than the traditional algorithms used in most parton shower programs currently, which may be of considerable importance for higher energy events or for the inclusion of more types of radiation.

\section*{Acknowledgement}
This work was supported by the Netherlands Foundation for Fundamental Research of Matter (FOM) programme entitled "Higgs as Probe and Portal".

\newpage
\bibliographystyle{JHEP}
\bibliography{mybib}

\section*{Appendix: Descriptions of the algorithms} \label{Appendix}
Here we give the algorithms described in the text. They are given in pseudocode and in unitary language.
\newpage
\renewcommand{\thealgorithm}{}
\begin{algorithm}[!]
\caption{The \emph{Veto-Max} full Sudakov veto algorithm}
\textbf{Input}
\begin{algorithmic}[1]
\STATE Branching kernels $p_i(t,z)$ with overestimates $q_i(t,z) = r_i(t)s_i(z)$ 
\STATE Boundaries $z_{i+}(t)$ and $z_{i-}(t)$ with overestimates $z_{i+}$ and $z_{i-}$ 
\STATE Integrated overestimate kernels $q_i(t)=r_i(t)\left(S_i(z_{i+}) - S_i(z_{i-}) \right)$ and their primitives $Q_i(t)$. 
\end{algorithmic}
\bigskip
\textbf{Algorithm}
\begin{algorithmic}
\STATE $t_{max} \leftarrow 0$
\FORALL {$1 \leq i \leq n$}
\STATE $t_i \leftarrow u$
\LOOP
\IF {$\rho_1 < e^{Q_i(\mu) - Q_i(u)}$}
\STATE $t_i \leftarrow \mu$
\STATE $z_i \leftarrow z_0$
\BREAK
\ELSE
\STATE $t_i \leftarrow Q_i^{-1}\left(\log\left(\rho_{1}\right)+
Q_i\left(t_i\right)\right)$
\IF {$t_i < t_{max}$}
\BREAK
\ENDIF
\STATE $z_i \leftarrow S_i^{-1}\left(\rho_{2}\left(S_i(z_{i+})-S_i(z_{i-})\right)+S_i(z_{i-})\right)$
\IF {$\rho_{3}<p_{i}(t_i,z_i)/q_{i}(t_i,z_i)$ and $z_{i-}(t) < z_i < z_{i+}(t)$}
\IF {$t_i > t_{max}$}
\STATE $t_{max} \leftarrow t_i$
\ENDIF
\BREAK
\ENDIF 
\ENDIF
\ENDLOOP
\ENDFOR
\STATE $j \leftarrow$ index(max$(t_i)$)
\RETURN $t_j$, $z_j$, $j$
\end{algorithmic}
\end{algorithm}

\begin{align}
E(t,z;u)&=\prod_{i=1}^{n} \biggl[ \biggr. \int dt_{i}\int dz_{j}\int_{0}^{u}d\tau_{i}\Bigl( \Bigr. q_{i}(\tau_{i})\exp\left(Q_{i}(\tau_{i})-Q_{i}(u)\right)\theta(\mu<\tau_{i}<u) \\ \nonumber
&+\exp(Q_{i}(\mu)-Q_{i}(u))\delta(\tau_{i}-\mu)\Bigl. \Bigr) \int_{z_{i-}}^{z_{i+}}d\zeta_{i}\frac{s_{i}(\zeta_{i})}{S_{i}(z_{i+})-S_{i}(z_{i-})} \\ \nonumber
&\times \Biggl\{ \Biggr. \theta(\tau_{i}=\mu)\delta(t_{i}-\mu)\delta(\zeta_i-z_0)+\theta(\tau_{i}\neq\mu) \biggl[ \biggr. (1-\theta_{i}^{\tau_{i}}(\zeta_{i}))E_{i}(t_{i},z_{i},\tau_{i}) \\ \nonumber
&+\theta_{i}^{\tau_{i}}(\zeta_{i})\int_{0}^{1}d\rho \biggl\{ \biggr. \theta\left(\rho<\frac{p_{i}(\tau_{i},\zeta_{i})}{q_{i}(\tau_{i},\zeta_{i})}\right)\delta(t_{i}-\tau_{i})\delta(z_{i}-\zeta_{i}) \\ \nonumber
&+\theta\left(\rho>\frac{p_{i}(\tau_{i},\zeta_{i})}{q_{i}(\tau_{i},\zeta_{i})}\right)E_{i}(t_{i},z_{i},\tau_{i}) \biggl. \biggr\} \biggl. \biggr] \Biggl. \Biggr\} \Biggl. \Biggr] \\ \nonumber
&\times \sum_{j=1}^{n}\theta(\max(t_{j}))\delta(t-t_{j})\delta(z-z_{j})
\end{align}

\newpage

\begin{algorithm}
\caption{The \emph{Max-Veto} full Sudakov veto algorithm}
\textbf{Input}
\begin{algorithmic}[1]
\STATE Branching kernels $p_i(t,z)$ with overestimates $q_i(t,z) = r_i(t)s_i(z)$ 
\STATE Boundaries $z_{i+}(t)$ and $z_{i-}(t)$ with overestimates $z_{i+}$ and $z_{i-}$ 
\STATE Integrated overestimate kernels $q_i(t)=r_i(t)\left(S_i(z_i+) - S_i(z_i-) \right)$ and their primitives $Q_i(t)$. 
\end{algorithmic}
\bigskip
\textbf{Algorithm}
\begin{algorithmic}
\STATE $t \leftarrow u$
\LOOP
\FORALL {$1 \leq i \leq n$}
\IF {$\rho_1 < e^{Q_i(\mu) - Q_i(u)}$}
\STATE $t_i \leftarrow \mu$
\ELSE 
\STATE $t_i \leftarrow Q_i^{-1}\left(\log\left(\rho_{1}\right)+
Q_i\left(t\right)\right)$
\STATE $z_i \leftarrow S_i^{-1}\left(\rho_{2}\left(S_i(z_{i+})-S_i(z_{i-})\right)+S_i(z_{i-})\right)$
\ENDIF
\ENDFOR
\IF {$t_j=\mu$}
\RETURN {$t_j, z_0, j$}
\ENDIF
\STATE $j \leftarrow$ index(max$(t_i)$)
\IF {$\rho_{3}<p_{j}(t_j,z_j)/q_{j}(t_j,z_j)$ and $z_{j-}(t) < z_j < z_{j+}(t)$}
\RETURN $t_j$, $z_j$, $j$
\ENDIF
\ENDLOOP
\end{algorithmic}
\end{algorithm}

\begin{align}
E(t,z;u)&=\prod_{i=1}^{n}\Biggl[ \Biggr. \int_{0}^{u}d\tau_{i}\Bigl( \Bigr. (q_{i}(\tau_{i})\exp(Q_{i}(\tau_{i})-Q_{i}(u))\theta(\mu<\tau_{i}<u) \\ \nonumber
&+\exp(Q_{i}(\mu)-Q_{i}(u))\delta(\tau_{i}-\mu) \Bigl. \Bigr) \int_{z_{i-}}^{z_{i+}}d\zeta_{i}\frac{s_{i}(\zeta_{i})}{S_{i}(z_{i+})-S_{i}(z_{i-})} \Biggl. \Biggr] \\ \nonumber
&\times \sum_{j=1}^{n}\theta(\max(\tau_{j})) \Biggl\{ \Biggr. \theta(\tau_{j}=\mu)\delta(t-\mu)\delta(\zeta_i-z_0) \\ \nonumber
&+\theta(\tau_{j}\neq\mu) \biggl[ \biggr. (1-\theta_{j}^{\tau_{j}}(\zeta_{j}))E(t,z,\tau_{j}) \\ \nonumber
&+\theta_{j}^{\tau_{j}}(\zeta_{j})\int_{0}^{1}d\rho \biggl\{ \biggr. \theta\left(\rho<\frac{p_{j}(\tau_{j},\zeta_{j})}{q_{j}(\tau_{j},\zeta_{j})}\right)\delta(t-\tau_{j})\delta(z-\zeta_{j}) \\ \nonumber
&+\theta\left(\rho<\frac{p_{j}(\tau_{j},\zeta_{j})}{q_{j}(\tau_{j},\zeta_{j})}\right)E(t,z,\tau_{j})\biggl. \biggr\} \biggl. \biggr] \Biggl. \Biggr\}
\end{align}

\newpage

\begin{algorithm}
\caption{The \emph{Generate-Select} Sudakov veto algorithm}
\textbf{Input}
\begin{algorithmic}[1]
\STATE Branching kernels $p_i(t,z)$ with overestimates $q_i(t,z) = r_i(t)s(z)$ 
\STATE Boundaries $z_{i+}(t)$ and $z_{i-}(t)$ with overestimates $z_{+}$ and $z_{-}$ 
\STATE Integrated overestimate kernels $q_i(t)=r_i(t)\left(S(z_+) - S(z_-) \right)$ and the primitive of their sum $\widetilde{Q}(t)$. 
\end{algorithmic}
\bigskip
\textbf{Algorithm}
\begin{algorithmic}
\STATE $t \leftarrow u$
\LOOP 
\IF{$\rho_{1} < e^{\widetilde{Q}(\mu) - \widetilde{Q}(u)}$}
\RETURN $\mu, z_0$
\ELSE
\STATE $t \leftarrow \widetilde{Q}^{-1}\left(\log\left(\rho_{1}\right)+
\widetilde{Q}\left(t\right)\right)$
\STATE $z \leftarrow S^{-1}\left(\rho_{2}\left(S(z_{+})-S(z_{-})\right)+S(z_{-})\right)$
\STATE Select $j$ between $1$ and $n$ with probability $q_{j}(t)/\widetilde{q}(t)$
\IF {$\rho_{3}<p_{j}(t,z)/q_{j}(t,z)$ and $z_{j-}(t) < z < z_{j+}(t')$}
\RETURN $t$, $z$, $j$
\ENDIF
\ENDIF
\ENDLOOP
\end{algorithmic}
\end{algorithm}

\begin{align}
E(t,z;u)&=\int_{0}^{u}d\tau\left(\widetilde{q}(\tau)\exp(\widetilde{Q}(\tau)-\widetilde{Q}(u))\theta(\mu<\tau<u)+\exp(\widetilde{Q}(\mu)-\widetilde{Q}(u))\delta(\tau-\mu)\right) \\ \nonumber
&\times \int_{z_{-}}^{z_{+}}d\zeta\frac{s(\zeta)}{S(z_{-})-S(z_{+})} \int_{0}^{1}d\rho\sum_{j=1}^{n}\theta\left(\frac{\sum_{i=1}^{j-1}q_{i}(\tau)}{\widetilde{q}(\tau)}<\rho<\frac{\sum_{i=1}^{j}q_{i}(\tau)}{\widetilde{q}(\tau)}\right) \\ \nonumber
&\times \Biggr[ \Biggl. \theta(\tau=\mu)\delta(t-\mu)\delta(z-z_0)+\theta(\tau\neq\mu) \Biggl\{ \Biggr. (1-\theta_{j}^{\tau}(\zeta))E(t,z,\tau) \\ \nonumber
&+\theta_{j}^{\tau}(\zeta)\int_{0}^{1}d\rho \biggl[ \biggr. \theta\left(\rho<\frac{p_{j}(\tau,\zeta)}{q_{j}(\tau,\zeta)}\right)\delta(t-\tau)\delta(z-\zeta) \\ \nonumber
&+\theta\left(\rho<\frac{p_{j}(\tau,\zeta)}{q_{j}(\tau,\zeta)}\right)E(t,z,\tau) \biggl. \biggr] \Biggl. \Biggr\} \Biggl. \Biggr]
\end{align}

\newpage

\begin{algorithm}
\caption{The \emph{Select-Generate} Sudakov veto algorithm}
\textbf{Input}
\begin{algorithmic}[1]
\STATE Branching kernels $p_i(t,z)$ with overestimates $q_i(t,z) = r_i(t)s_i(z)$ 
\STATE Boundaries $z_{i+}(t)$ and $z_{i-}(t)$ with overestimates $z_{i+}$ and $z_{i-}$ 
\STATE Integrated overestimate kernels $q_i(t)=r_i(t)\left(S_i(z_{i+}) - S(z_{i-}) \right)$, all with the same $t$-dependence, and the primitive of their sum $\widetilde{Q}(t)$. 
\end{algorithmic}
\bigskip
\textbf{Algorithm}
\begin{algorithmic}
\STATE $t \leftarrow u$
\LOOP 
\IF{$\rho_{1} < e^{\widetilde{Q}(\mu) - \widetilde{Q}(u)}$}
\RETURN $\mu, z_0$
\ELSE
\STATE Select $j$ between $1$ and $n$ with probability $q_{j}(t)/\widetilde{q}(t)$
\STATE $t \leftarrow \widetilde{Q}^{-1}\left(\log\left(\rho_{1}\right)+
\widetilde{Q}\left(t\right)\right)$
\STATE $z \leftarrow S_j^{-1}\left(\rho_{2}\left(S_j(z_{j+})-S_j(z_{j-})\right)+S_j(z_{j-})\right)$
\IF {$\rho_{3}<p_{j}(t,z)/q_{j}(t,z)$ and $z_{j-}(t) < z < z_{j+}(t')$}
\RETURN $t$, $z$, $j$
\ENDIF
\ENDIF
\ENDLOOP
\end{algorithmic}
\end{algorithm}

\begin{align}
E(t,z;u)&=\int_{0}^{1}d\rho\sum_{j}\theta\left(\frac{\sum_{i=0}^{j-1}q_{i}(u)}{\widetilde{q}(u)}<\rho<\frac{\sum_{i=0}^{j}q_{i}(u)}{\widetilde{q}(u)}\right) \\ \nonumber
&\times\int_{0}^{u}d\tau\left(\widetilde{q}(\tau)\exp(\widetilde{Q}(\tau)-\widetilde{Q}(u))\theta(\mu<\tau<u)+\exp(\widetilde{Q}(\mu)-\widetilde{Q}(u))\delta(\tau-\mu)\right) \\ \nonumber
&\times\int_{z_{j-}}^{z_{j+}}d\zeta\frac{s_{j}(\zeta)}{S_{j}(z_{j-})-S_{j}(z_{j+})} \\ \nonumber
&\times \Biggr[ \Biggl. \theta(\tau=\mu)\delta(t-\mu)\delta(z-z_0)+\theta(\tau\neq\mu) \Biggl\{ \Biggr. (1-\theta_{j}^{\tau}(\zeta))E(t,z,\tau) \\ \nonumber
&+\theta_{j}^{\tau}(\zeta)\int_{0}^{1}d\rho \biggl[ \biggr. \theta\left(\rho<\frac{p_{j}(\tau,\zeta)}{q_{j}(\tau,\zeta)}\right)\delta(t-\tau)\delta(z-\zeta) \\ \nonumber
&+\theta\left(\rho<\frac{p_{j}(\tau,\zeta)}{q_{j}(\tau,\zeta)}\right)E(t,z,\tau) \biggl. \biggr] \Biggl. \Biggr\} \Biggl. \Biggr]
\end{align}

\end{document}